\documentclass[twocolumn,showpacs,amsmath,amssymb]{revtex4}
\usepackage{graphicx,color}

\begin{document}

\title{Predator-prey cycles from resonant amplification of demographic 
stochasticity}

\author{A. J. McKane}
\affiliation{Theory Group, School of Physics and Astronomy, University of 
Manchester, Manchester M13 9PL, UK}
\author{T. J. Newman}
\affiliation{Department of Physics and Astronomy, Arizona State University,
Tempe, AZ 85287, USA\\
School of Life Sciences, Arizona State University, Tempe, AZ 85287, USA}
 
\begin{abstract}
In this paper we present the simplest individual level model of
predator-prey dynamics and show, via direct calculation, that it
exhibits cycling behavior. The deterministic analogue of our model, 
recovered when the number of individuals is infinitely large, is the 
Volterra system (with density-dependent prey reproduction) 
which is well-known to fail to predict cycles. This 
difference in behavior can be traced to a resonant amplification 
of demographic fluctuations which disappears only when the number of 
individuals is strictly infinite. Our results indicate that 
additional biological mechanisms, such as predator satiation, may
not be necessary to explain observed predator-prey cycles in real (finite) 
populations. 
\end{abstract} 


\maketitle
 
Predator-prey cycles are one of the most striking phenomena observed
in population biology, and as such, inspire intense discussion among
ecologists~\cite{ber02,tur03}. Cycles are also seen in a wide variety of 
other ``host-natural enemy'' systems, such as host-pathogen~\cite{and86} 
systems -- one of the most well known examples is measles, epidemics 
of which have been studied for many years~\cite{and91}. In this 
paper, we will be concerned with modeling the phenomenon of cycles, and 
will focus on predator-prey systems, for concreteness, but our main results 
will have direct applicability to other host-natural enemy systems, since 
they can be modeled in a similar way, often using identical equations. We
also believe that, since the phenomenon we describe is quite generic in 
certain classes of stochastic systems, it should be found outside population
dynamics. It seems that the precise mechanism underlying the 
existence of the cycles has not so far been elucidated because it 
involves the analysis of stochastic systems with a large, but finite, number 
of constituents, and also because it involves concepts such as resonance,
which are more familiar to physicists than biologists.
 
Among the numerous hypotheses put forward to explain cycles, perhaps
the simplest is that cycles arise directly from predator-prey
interactions.  Within this conceptual framework, theoretical modeling
of cycles has traditionally been developed using deterministic
population-level models (PLMs). Such discussions begin with
Volterra-like equations, which are coupled differential equations for 
the predator and prey densities.  Equations of this type
encapsulate the simplest processes of predator and prey mortality,
prey reproduction and competition, and predation.  Surprisingly these
models do not predict stable cycles: additional biological mechanisms,
such as predator satiation,
need to be included within the framework of differential equations to
give cycles~\cite{may74}. It seems puzzling that cycles, 
which are so easy to understand intuitively, can only be described 
mathematically in models which include these more subtle mechanisms. 
In order to probe this issue, we shall take a different approach here, 
and describe the predator-prey system using an individual level model 
(ILM). The individuals, which are either predators or prey, are acted 
upon by simple stochastic processes of mortality, reproduction, and 
predation. We are able to derive an exact description of this model 
when the number of individuals is large and finite.  We find that the 
predator and prey numbers undergo large cycles, just as one would 
expect intuitively. The cycles, which arise from a novel resonance 
effect, disappear only when the number of individuals is taken to be 
strictly infinite, that is, when the PLM is recovered. From a statistical
physics viewpoint, we would term the PLM a mean field theory of the underlying
``microscopic'' ILM which includes statistical fluctuations.

Predator-prey cycles observed in nature will have a stochastic 
component --- this will affect both their amplitude and phase. Therefore 
care must be taken in averaging over replicates. A direct average of the 
population densities from different replicates will result in a constant 
average density since, in the absence of an external ``forcing'',
there is a lack of synchrony between the cycles from different replicates.
This fact is crucial when modeling predator-prey cycles. A given PLM is 
written in terms of a population density, which can be thought of as the 
result of an average of the population numbers from a large number of ILM 
replicates. If a given ILM shows oscillatory behavior, such cycles will be
lost in the modeling transition to a PLM. Thus, it is necessary
to study quantities such as the autocorrelation function and power 
spectrum arising from an ensemble of ILMs, in order to determine the 
presence and properties of predator-prey cycles.

The specific ILM we study in this paper is a non-spatial
stochastic model. At a given time, a realization of the
ILM consists of $n$ individuals
of species $A$ (the predators) and $m$ individuals of species $B$ (the
prey). Since we are interested in what is essentially the simplest model
of predator-prey interactions, we 
include only birth processes $BE \stackrel {b}{\rightarrow } BB$, 
death processes $A \stackrel {d_{1}}{\rightarrow } E$, $B \stackrel
{d_{2}}{\rightarrow } E$, and predator-prey interactions
$AB \stackrel {p_{1}} {\rightarrow } AA$, $AB \stackrel {p_{2}}{\rightarrow }
AE$. Here $(b, d_{1},d_{2}, p_{1}, p_{2})$ are rate constants. 
The symbol $E$ corresponds to what would be available sites 
in a spatial model. In this non-spatial model, the $E$'s 
are $(N-n-m)$ passive constituents of the system, which are required
for prey reproduction, and which result in intra-specific prey competition.
Note, the overall number of $A, B$ and $E$ constituents is fixed to be $N$. 
The dynamics of the model consists of choosing 
constituents at random and implementing the rules given above. The time
dynamics of the model can either be numerically simulated or studied
analytically using the formalism of master equations~\cite{ren91,van92}. 
In the latter case the transition rates $T(n',m'|n,m)$ from the state $(n,m)$ 
to the state $(n',m')$ are given by
\begin{eqnarray}
T(n-1,m|n,m) &=& d_{1} n\,, \nonumber \\
T(n,m+1|n,m) &=& 2 b \frac{m}{N} (N - n - m)\,, \nonumber \\ 
T(n,m-1|n,m) &=& 2 p_{1} \frac{n m}{N} + d_{2} m\,, \nonumber \\
T(n+1,m-1|n,m) &=& 2 p_{2} \frac{n m}{N}\,,
\label{trans_rates}
\end{eqnarray}
where the $b$ and $d_i$ have been scaled by a factor of $(N-1)$ and the
$d_i$ by a factor of $N$.

We stress that the individuals of a given species in our model are 
identical, and thus the term ILM should not be confused with ``agent based 
models'' which are often designed to study the ecological effects of 
behavioral and physiological variation among individuals. We have already 
given an extensive discussion of this approach elsewhere in the context 
of competition models~\cite{mck04}, and we refer the reader to this paper 
for a fuller discussion of the formalism. For predator-prey models however, 
including stochastic dynamics leads to more marked effects than in competition 
models, as we now discuss.

The master equation for the probability that the system consists of $n$ 
predators and $m$ prey at time $t$, $P(n,m,t)$, is
\begin{eqnarray}
\frac{dP(n,m,t)}{dt} &=& ( {\cal E}_{x} - 1 )\, 
\left[ T(n-1,m|n,m)P(n,m,t) \right] \nonumber \\
+ ( {\cal E}^{-1}_{y} &-& 1 )\, 
\left[ T(n,m+1|n,m)P(n,m,t) \right] \nonumber \\
+ ( {\cal E}_{y} &-& 1 )\, 
\left[ T(n,m-1|n,m)P(n,m,t) \right] \nonumber \\
+ ( {\cal E}^{-1}_{x} {\cal E}_{y} &-& 1 )\, 
\left[ T(n+1,m-1|n,m)P(n,m,t) \right]\,,
\label{master}
\end{eqnarray}
where the step operators $\cal E$ are defined by their actions on functions
of $n$ and $m$ by ${\cal E}^{\pm 1}_{x}f(n,m,t) = f(n \pm 1,m,t)$ and 
${\cal E}^{\pm 1}_{y}f(n,m,t) = f(n,m \pm 1,t)$. 

The mean field limit of this ILM may be obtained by multiplying (\ref{master})
by $n$ and $m$ in turn, and subsequently summing over all allowed values of
$m$ and $n$. This gives equations for the mean values 
$f_{1}=\langle n \rangle/N$ and $f_{2}=\langle m \rangle/N$ in the limit 
$N \to \infty$ if we ignore terms which are $1/N$ down on others and make the
replacements $\langle m^{2} \rangle \rightarrow \langle m \rangle^{2}$ and
$\langle m n \rangle \rightarrow \langle m \rangle \langle n \rangle$. This
mean field theory, or PLM, takes the form
\begin{eqnarray}
\frac{df_1}{dt} &=& n(f_{2})f_{1} - \mu f_{1} \nonumber \\
\frac{df_2}{dt} &=& rf_{2}\left( 1 - \frac{f_{2}}{K} \right) - g(f_{2})f_{1}\,.
\label{Volterra}
\end{eqnarray}
The eqs. (\ref{Volterra}) 
are frequently referred to as the Volterra equations, to distinguish them
from the Lotka-Volterra equations which have no term in 
$f_{2}/K$~\cite{ren91}. The constants $\mu$, $r$, and $K$ are simply functions 
of the rate constants:
\begin{equation}
\mu = d_{1}\,, \ \ r = 2b-d_{2}\,, \ \ K = 1 - \frac{d_{2}}{2b}\,,
\label{const_rateconst}
\end{equation}
and the linear numerical and functional responses are given by
$n(f_{2})=2p_{1}f_{2}$ and $g(f_{2})=2(p_{1}+p_{2}+b)f_{2}$ respectively. 


\begin{figure}[b!]
\begin{center}
\includegraphics[width=7.5cm]{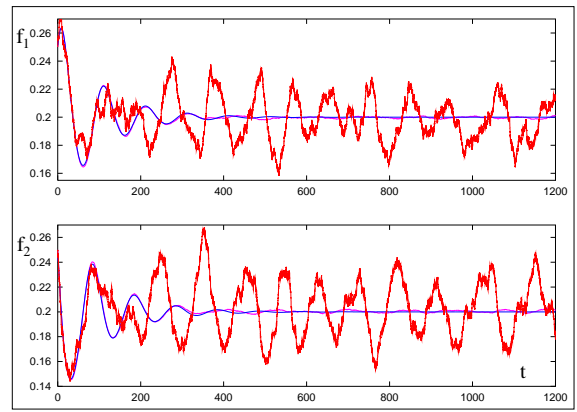}
\caption{Predator and prey densities as a function of time.
The upper panel shows the predator density $f_{1}$
for $N=3200$. The blue line is calculated from numerical integration
of the mean field Volterra Eqs.~(\ref{Volterra}). The purple line is the
average of the predator density time series from 500 replicates 
generated from the ILM, and is almost indistinguishable from the mean 
field solution. The red line is the predator density time series for
a single typical replicate. The lower panel is the equivalent plot
for the prey density $f_{2}$. Parameter values are $b=0.5$, $d_{1}=0.1$,
$d_{2}=0.0$, $p_{1}=0.5$, and $p_{2}=0.1$.}
\label{fig1}
\end{center}
\end{figure}


As is well known~\cite{ren91}, the analysis of this model shows a
complete absence of cycles. There is a single
fixed point for which the predators and prey have non-zero population sizes.
Denoting these stationary values by $f^{(s)}_{1}$ and $f^{(s)}_{2}$, then in
terms of the original rate constants they are given by:
\begin{equation}
\label{ss}
f^{(s)}_{1}=
\frac{(2bp_{1}-bd_{1}-p_{1}d_{2})}{2p_{1}(p_{1}+p_{2}+b)},\
f^{(s)}_{2} = \frac{d_{1}}{2p_{1}}\,.
\end{equation}
The stability of this fixed point may be studied may performing linear
stability analysis. This results in a stability matrix which is given by
\begin{equation} 
A = \left(
\begin{array}{cc}
0 & 2p_{1} f^{(s)}_{1} \\
-2 (p_{1}+p_{2}+b) f^{(s)}_{2} & -2b f^{(s)}_{2}
\end{array}
\right)\,.
\label{stability}
\end{equation} 
We have expressed the entries in terms of the fixed point values, since these 
are manifestly positive, and it is easy to see that the eigenvalues of $A$ 
both have a negative real part, implying that the fixed point is stable. The
entries of $A$ will appear again below 
in the analysis of the cycling behavior for finite $N$. For 
now let us simply remark that, while there is no limit cycle in the Volterra 
system (\ref{Volterra}), a limit cycle does exist in the Lotka-Volterra
equations (obtained by taking $K \rightarrow \infty$), 
but it is neutrally stable due to a conserved quantity in 
the model. This unrealistic behavior disappears with the introduction of
a finite carrying capacity, $K$, in (\ref{Volterra}), but, as mentioned
above, leads to a complete absence of cycling behavior. 
We will show below that cycles can be found
in the ILM, but only when $N$ is finite; the $N \to \infty$ limit which was
taken in order to derive the PLM, eliminates the cycles present in the
original ILM.

To see this, let us first note that the ensemble averaged population density 
of the ILM, determined from numerical simulations, agrees beautifully with the 
solution of this deterministic model (Fig.~1, purple and blue lines 
respectively) showing a decaying oscillatory transient followed by a constant 
steady-state density, typical of a Volterra system. In marked contrast, 
individual realizations of the ILM show large persistent cycles (Fig.~1, red 
line). The amplitude of these oscillations is much larger than the naive 
estimate based on the law of large numbers. In fact, the oscillations are of 
order $(1/\sqrt{N})$ as would be expected, but amplified by a very large 
factor due to a noise-induced resonance effect, as explained below. 

This cycling behavior can be investigated analytically by extracting an 
``effective theory'' valid for large $N$, through applying a standard method 
to the master equations, due to van Kampen~\cite{van92}. Essentially the 
method involves the replacements 
$n/N = f_{1} + x/\sqrt{N}$ and $m/N = f_{2} + y/\sqrt{N}$ in the transition
probabilities that appear in the master equation. By changing from a 
description based on the (discrete) variables $n$ and $m$ to one based on the
(continuous) variables $x$ and $y$, terms of different orders in $1/N$ can
be identified in the master equation: the leading order terms gives rise to
a deterministic set of equations and the next-to-leading order terms give rise
to a linear Fokker-Planck equation. The leading order set of equations (mean 
field theory) are the PLM and are the Volterra equations (\ref{Volterra}) 
which we have already obtained by a more direct method.

At next-to-leading order, rather than write down the Fokker-Planck equation, 
it is simpler to give the set of Langevin equations to which it is 
equivalent~\cite{van92}. They take the form
\begin{eqnarray}
\dot{x} &=& a_{11}x+a_{12}y+\eta_{1}(t) \nonumber \\
\dot{y} &=& a_{21}x+a_{22}y+\eta_{2}(t)\,.
\label{Langevin}
\end{eqnarray}
These are a pair of differential equations which describe the stochastic 
behavior of the ILM at large $N$: $x(t)$ and $y(t)$ are stochastic 
corrections to the deterministic behavior of the predator and prey
densities respectively, at large but finite $N$. The constants, $a_{ij}$, 
appearing in Eq.(\ref{Langevin}) are exactly the entries of the matrix $A$,
Eq.~(\ref{stability}) found from a linear stability analysis
about the non-trivial fixed point of Eq.(\ref{Volterra}). The noise
covariance matrix $b_{ij}$, 
which is responsible for generating the large-scale
oscillations, cannot be determined from Eq.(\ref{Volterra}) and is
derived from the master equation using the van Kampen expansion. Since
the noise is white, $b_{ij} = \langle \tilde{\eta}_{i} (\omega)
\tilde{\eta}_{j} (-\omega) \rangle$ is independent of the frequency
$\omega$.  The explicit expressions for these constants are
\begin{eqnarray}
\nonumber b_{11} & = & 2d_{1}f^{(s)}_{1},\\ 
b_{12} & = & b_{21}= - d_{1} f^{(s)}_{1}, \nonumber \\ 
b_{22} & = & 2d_{1}(1+p_{2}/p_{1})f^{(s)}_{1} + 2d_{2} f^{(s)}_{2}\,.
\label{bij}
\end{eqnarray}
As discussed earlier, it is not the average behavior of replicates that 
interests us, but rather measures which characterize the oscillations. 
Examining $x$ and $y$ as functions of frequency allow us to determine the 
nature of the oscillations.


\begin{figure}[b!]
\begin{center}
\includegraphics[width=7.5cm]{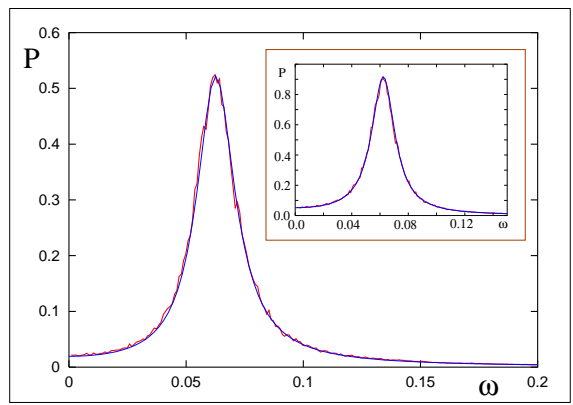}
\caption{A plot of the power spectrum $P(\omega )$ for
the predator time series, as a 
function of frequency $\omega $. The red line corresponds to $P$
calculated from 500 replicate runs of the ILM. The blue line is
the prediction from our theory, namely Eq.~(\ref{power}). 
The parameter values are the same as those described
in the caption to Fig.~1. The inset shows the analogous power spectra
(data and theory) for the prey time series.}
\label{fig2}
\end{center}
\end{figure}


To search for oscillations in noisy data, one of the most useful
diagnostic tools is the power spectrum $P(\omega ) = \langle
|\tilde{x}(\omega)|^{2} \rangle$, where $\tilde{x}(\omega)$ is the
Fourier transform of $x (t)$. Taking the Fourier transform of 
Eq.~(\ref{Langevin}), solving for ${\tilde x}(\omega )$, and averaging 
its squared modulus, we find
\begin{equation}
P(\omega ) = \frac {\alpha + \beta \omega ^{2}} 
{[(\omega ^{2}-\Omega_{0}^{2})^{2}+\Gamma ^{2}\omega ^{2}]}\,,
\label{power}
\end{equation}
where $\alpha$ and $\beta$ are functions of the ILM rates:
$\alpha = b_{11} a_{22}^{2} + 2 b_{12} a_{12} |a_{22}| + b_{22} a_{12}^{2}$ 
and $\beta = b_{11}$. The constants in the denominator have the especially 
simple forms: $\Omega^{2}_{0}=a_{12} |a_{21}|$ and $\Gamma = |a_{22}|$. The 
spectrum predicted by Eq.~(\ref{power}) gives the blue line shown in Fig.~2. 
The agreement with the spectrum obtained from simulation of the ILM (red line) 
is excellent. Note, the naive $O(1/\sqrt{N})$ estimate of the size of 
stochastic fluctuations corresponds to the zero frequency value of 
$P(\omega)$. Fig.~2 clearly illustrates the very large amplification
of these fluctuations due to the resonance effect.

The spectrum given above is reminiscent of that for a simple
mechanical system --- namely a linear damped
harmonic oscillator, with natural frequency $\Omega _{0}$ and
driven at frequency $\omega $. In a mechanical oscillator the
driving frequency must be tuned to achieve resonance. In the stochastic
predator-prey model described here no tuning is necessary. The system
is driven by white noise, as shown in Eq.~(\ref{Langevin}), which 
covers all frequencies --- thus the resonant frequency of the system is
excited without tuning. We stress, the noise which drives the
system is {\it not external}, but arises from the demographic stochasticity
contained in the individual processes which define the model. 
We also stress that there is no external or environmental
stochasticity in our model, and that the resonance phenomenon we
report here is not related to ``stochastic resonance''. 

The damping term, represented by the constant $\Gamma $, limits the amplitude
of the oscillations. Predator-prey systems for which $\Gamma $ happens
to be small will be at risk of extinction through resonant oscillations,
despite having large population sizes. The resonant oscillation occurs in 
the regime $2 a_{12} |a_{21}| > a_{22}^{2}$, where the
resonant frequency $\omega_{0} = \sqrt{\Omega^{2}_{0} - \Gamma^{2}/2}$
is real. A similar analysis can be carried out to obtain the 
spectrum for the prey time series.  It again has the form Eq.~(\ref{power}), 
but now with $\alpha = b_{11} a_{21}^{2}$ and $\beta = b_{22}$. The
positions of the peaks for these two power spectra are only weakly
dependent on the $\alpha$'s and $\beta$'s, and so they are almost
coincident.

Predator-prey systems (and related host-pathogen systems) have been
studied theoretically for decades. Most of the previous studies have focused 
on the role of environmental stochasticity, the relevance of non-linear 
interactions or of spatial effects, to explain the mechanism of 
cycling~\cite{nis82,ren91,kai96,apa01,bjo01,pas01,pas03}. Some authors have 
discussed the role that demographic stochasticity may have on  
cycles~\cite{bar60,ren91}. Most of this discussion has been qualitative; the
nearest to our own discussion was a prescient analysis by Bartlett nearly
fifty years ago~\cite{bar60}, in which he postulated equations similar to 
(\ref{Langevin}). However, he did not note the existence of a resonance, and 
so proposed cycles with an amplitude which were not enhanced by this effect, 
and which were therefore of limited biological interest.

The idea that external perturbations with a dominant frequency can 
entrain the predator-prey dynamics in a cyclic nature is fairly intuitive;
the phenomenon we discuss here is more fundamental and less intuitive. The 
noise in our system is internal --- purely a result of the demographic 
stochasticity inherent in discrete birth, death, and predation events. The 
key point is that this internal noise has a flat power spectrum in the 
frequency domain; in other words, it excites all frequencies of the system 
simultaneously. These excitations are typically of limited interest in a 
large population of $N$ individuals, since they give rise to small 
$O(1/\sqrt{N})$ fluctuations about the mean population densities.
Such is the case, for example, in competition models~\cite{mck04}. 
The predator-prey system, and related ones such as epidemic models, are 
exceptional, in that the equations describing linear fluctuations about the 
steady-state are susceptible to resonant amplification in the vicinity of an 
internal frequency $\Omega $, which is a property of the population itself. 
The internal noise, in exciting all frequencies, automatically resonates the
system giving rise to large oscillations in the population densities.
This phenomenon is ``emergent'' in the truest sense of the word. We expect
that this resonance mechanism will occur in other stochastic 
systems in which the mean field theory shows damped oscillations. 

\acknowledgments 
We thank D. Alonso, J. Antonovics, M. Chubynsky, M. Pascual and J. 
Vandermeer for useful discussions. We acknowledge the NSF for partial 
support, under grant DEB-0328267.


\begin{references} 

\bibitem{ber02} A.~A. Berryman, \emph{Population Cycles}. (Oxford
University Press, Oxford, 2002).

\bibitem{tur03} P. Turchin, \emph{Complex Population Dynamics}. (Princeton
University Press, Princeton, 2003).

\bibitem{and86} R.~M. Anderson and R.~M. May, Phil. Trans. Roy. Soc. B 
\textbf{314}, 533 (1986).

\bibitem{and91} R.~M. Anderson and R.~M. May, \emph{Infectious Diseases 
of Humans}. (Oxford University Press, Oxford, 1991).

\bibitem{may74} J. Maynard Smith, \emph{Models in Ecology}. 
(Cambridge University Press, Cambridge, 1974).

\bibitem{ren91} E. Renshaw, \emph{Modelling biological populations 
in space and time}. (Cambridge University Press, Cambridge, 1991).

\bibitem{van92} N.~G. van Kampen, \emph{Stochastic Processes in Physics
and Chemistry}. (Elsevier, Amsterdam, 1992).

\bibitem{mck04} A.~J. McKane and T.~J. Newman, Phys. Rev E \textbf{70},
041902 (2004).

\bibitem{nis82} R. Nisbet and W. Gurney, \emph{Modelling Fluctuating
Populations}. (Wiley, New York, 1982).

\bibitem{kai96} V. Kaitala, E. Ranta and J. Lindstrom, J. Anim. Ecol. 
\textbf{65}, 249 (1996).

\bibitem{apa01} J.~P. Aparicio and H.~G. Solari, Math. Bioscience 
\textbf{169}, 15 (2001).

\bibitem{bjo01} O.~N. Bj{\o}rnstad and B.~T. Grenfell, Science \textbf{293}, 
638 (2001).

\bibitem{pas01} M. Pascual, P. Mazzaga and S.~A. Levin, Ecology \textbf{82}, 
2357 (2001).

\bibitem{pas03} M. Pascual and P. Mazzega, Theor. Popul. Biol. \textbf{64}, 
385 (2003).

\bibitem{bar60} M.~S. Bartlett, \textit{Stochastic Population Models}
(Methuen, London, 1960), p\,64.  

\end{references}
\end{document}